\documentclass[%
reprint,
superscriptaddress,
nofootinbib,
nobibnotes,
 amsmath,
 amssymb,
longbibliography,
]{revtex4-2}

\usepackage{amsmath}
\usepackage{wasysym}
\usepackage{braket}

\usepackage{gensymb}
\usepackage{float}
\usepackage{graphicx}
\usepackage{dcolumn}
\usepackage{bm}
\usepackage{upgreek} 
\usepackage{multirow} 

\usepackage{mathrsfs}

\makeatletter
\renewcommand\frontmatter@abstractwidth{\dimexpr\textwidth\relax}
\renewcommand{\@biblabel}[1]{\quad#1.}
\makeatother

\usepackage{multibib}
\newcites{paper}{References}

\setcitestyle{numbers,super,open={},close={}}

\begin{document}

\title{Entangling single atoms over 33 km telecom fibre}

\author{Tim van Leent}
\altaffiliation[ ]{These authors contributed equally}
\affiliation{Fakult{\"a}t f{\"u}r Physik, Ludwig-Maximilians-Universit{\"a}t M{\"u}nchen, Schellingstr. 4, 80799 M{\"u}nchen, Germany}
\affiliation{Munich Center for Quantum Science and Technology (MCQST), Schellingstr. 4, 80799 M{\"u}nchen, Germany}

\author{Matthias Bock}
\altaffiliation[ ]{These authors contributed equally}
\affiliation{Fachrichtung Physik, Universit{\"a}t des Saarlandes, Campus E2.6, 66123 Saarbr{\"u}cken, Germany}

\author{Florian Fertig}
\altaffiliation[ ]{These authors contributed equally}
\affiliation{Fakult{\"a}t f{\"u}r Physik, Ludwig-Maximilians-Universit{\"a}t M{\"u}nchen, Schellingstr. 4, 80799 M{\"u}nchen, Germany}
\affiliation{Munich Center for Quantum Science and Technology (MCQST), Schellingstr. 4, 80799 M{\"u}nchen, Germany}

\author{Robert Garthoff}
\affiliation{Fakult{\"a}t f{\"u}r Physik, Ludwig-Maximilians-Universit{\"a}t M{\"u}nchen, Schellingstr. 4, 80799 M{\"u}nchen, Germany}
\affiliation{Munich Center for Quantum Science and Technology (MCQST), Schellingstr. 4, 80799 M{\"u}nchen, Germany}

\author{Sebastian Eppelt}
\affiliation{Fakult{\"a}t f{\"u}r Physik, Ludwig-Maximilians-Universit{\"a}t M{\"u}nchen, Schellingstr. 4, 80799 M{\"u}nchen, Germany}
\affiliation{Munich Center for Quantum Science and Technology (MCQST), Schellingstr. 4, 80799 M{\"u}nchen, Germany}

\author{Yiru Zhou}
\affiliation{Fakult{\"a}t f{\"u}r Physik, Ludwig-Maximilians-Universit{\"a}t M{\"u}nchen, Schellingstr. 4, 80799 M{\"u}nchen, Germany}
\affiliation{Munich Center for Quantum Science and Technology (MCQST), Schellingstr. 4, 80799 M{\"u}nchen, Germany}

\author{Pooja Malik}
\affiliation{Fakult{\"a}t f{\"u}r Physik, Ludwig-Maximilians-Universit{\"a}t M{\"u}nchen, Schellingstr. 4, 80799 M{\"u}nchen, Germany}
\affiliation{Munich Center for Quantum Science and Technology (MCQST), Schellingstr. 4, 80799 M{\"u}nchen, Germany}

\author{Matthias Seubert}
\affiliation{Fakult{\"a}t f{\"u}r Physik, Ludwig-Maximilians-Universit{\"a}t M{\"u}nchen, Schellingstr. 4, 80799 M{\"u}nchen, Germany}
\affiliation{Munich Center for Quantum Science and Technology (MCQST), Schellingstr. 4, 80799 M{\"u}nchen, Germany}

\author{Tobias Bauer}
\affiliation{Fachrichtung Physik, Universit{\"a}t des Saarlandes, Campus E2.6, 66123 Saarbr{\"u}cken, Germany}

\author{Wenjamin Rosenfeld}
\affiliation{Fakult{\"a}t f{\"u}r Physik, Ludwig-Maximilians-Universit{\"a}t M{\"u}nchen, Schellingstr. 4, 80799 M{\"u}nchen, Germany}
\affiliation{Munich Center for Quantum Science and Technology (MCQST), Schellingstr. 4, 80799 M{\"u}nchen, Germany}

\author{Wei Zhang}
\altaffiliation{Email: wei.zhang@xjtu.edu.cn}
\altaffiliation{Present address: Xi'An Jiao Tong University, Xi'An ShannXi, China}
\affiliation{Fakult{\"a}t f{\"u}r Physik, Ludwig-Maximilians-Universit{\"a}t M{\"u}nchen, Schellingstr. 4, 80799 M{\"u}nchen, Germany}
\affiliation{Munich Center for Quantum Science and Technology (MCQST), Schellingstr. 4, 80799 M{\"u}nchen, Germany}

\author{Christoph Becher}
\altaffiliation{Email: christoph.becher@physik.uni-saarland.de}
\affiliation{Fachrichtung Physik, Universit{\"a}t des Saarlandes, Campus E2.6, 66123 Saarbr{\"u}cken, Germany}
\email{christoph.becher@physik.uni-saarland.de}

\author{Harald Weinfurter}
\altaffiliation{Email: h.w@lmu.de}
\affiliation{Fakult{\"a}t f{\"u}r Physik, Ludwig-Maximilians-Universit{\"a}t M{\"u}nchen, Schellingstr. 4, 80799 M{\"u}nchen, Germany}
\affiliation{Munich Center for Quantum Science and Technology (MCQST), Schellingstr. 4, 80799 M{\"u}nchen, Germany}
\affiliation{Max-Planck Institut f{\"u}r Quantenoptik, Hans-Kopfermann-Str. 1, 85748 Garching, Germany}

\date{\today}

\begin{abstract}
\noindent \textbf{Heralded entanglement between distant quantum memories is the key resource for quantum networks. Based on quantum repeater protocols, these networks will facilitate efficient large-scale quantum communication and distributed quantum computing. However, despite vast efforts, long-distance fibre based network links have not been realized yet. Here we present results demonstrating heralded entanglement between two independent, remote single-atom quantum memories generated over fibre links with a total length up to 33 km. To overcome the attenuation losses in the long optical fibres of photons initially emitted by the Rubidium quantum memories, we employ polarization-preserving quantum frequency conversion to the low loss telecom band. The presented work represents a milestone towards the realization of efficient quantum network links.}
\end{abstract}

\maketitle

\noindent Sharing entanglement between distant quantum systems is a crucial ingredient for the realization of quantum networks \citepaper{briegel1998quantum, kimble2008quantum}. Photons are the tool of choice to mediate entanglement distribution, typically either via controlled light-matter interaction with local memories \citepaper{brekenfeld2020quantum,daiss2021quantum}, or, as it also will be used here, via entanglement swapping from two pairs of entangled photon-memory states \citepaper{zukowski1993event,hofmann2012heralded,pompili2021realization,liu2021heralded}. Innovative applications of such networks include distributed quantum computing \citepaper{cuomo2020towards} and device-independent quantum key distribution \citepaper{acin2007device}. Since attenuation losses in the distribution process are inevitable, quantum repeaters will be essential to efficiently distribute entanglement via intermediate nodes.

To minimize absorption along the quantum channel and thus to maximize the distance between neighbouring nodes in quantum networks employing the readily available fibre infrastructure, it is necessary to convert light to telecom wavelengths~\citepaper{zaske2012visible, ates2012two, de2012quantum, ikuta2011wide, maring2017photonic}. Light-matter entanglement distributed at the low loss telecom band has recently been demonstrated for various types of quantum memories, including NV-centres, ions, atoms, and atomic ensembles \citepaper{bock2018high,ikuta2018polarization,dudin2010entanglement,Tchebotareva2019}, even over tens of kilometres of fibre \citepaper{krutyanskiy2019light,leent2020long}. This was mainly enabled by novel quantum frequency converters \citepaper{bock2018high,ikuta2018polarization}, which, while preserving the photonic polarization, have reached external device conversion efficiencies as high as 57\%~\citepaper{leent2020long}.
 
For future quantum communication and repeater scenarios, it is vital that the nodes are independent and distant, employ long-lived quantum memories, and, at the same time, provide the availability of heralded entanglement, i.e., there is a well defined signal available that the entanglement distribution succeeded. So far, this has been limited to fibre lengths up to 1.7 km \citepaper{hensen2015loophole,rosenfeld2017event}. Recently, great progress was made by demonstrating telecom-heralded entanglement between atomic ensembles \citepaper{yu2020entanglement} and multimode solid-state quantum memories \citepaper{lago2021telecom}, however, having limited memory storage times and not employing independent nodes.

Here we report on the distribution of entanglement between two remote quantum nodes---$^{87}$Rb atoms trapped and manipulated independently at locations 400 m apart---generated over fibre links with a length of up to 33 km. The scheme begins with entangling the spin state of an atom with the polarization state of a photon in each node. Subsequently, the photons emitted by the atoms at 780 nm are converted to telecom wavelength and transferred over several kilometres of fibre to a middle station, where a Bell-state measurement is performed to swap the entanglement to the atoms. We analyse the heralded entanglement between the atoms for different fibre link lengths using correlation measurements of the atom-atom state along three bases. The atoms are analysed after a delay allowing for two-way communication to the middle station over the full fibre link length to realistically evaluate the performance for long fibre lengths. \\

\begin{figure*}
\includegraphics[width=0.9\linewidth]{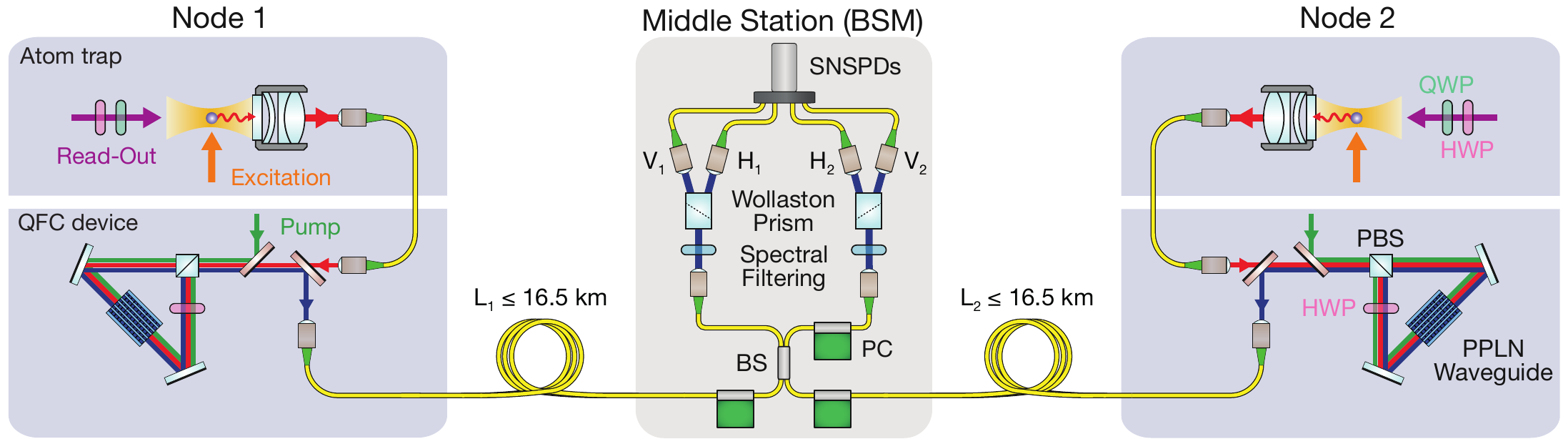}
\caption{\protect\textbf{Schematic of the experimental setup.} 
In each node, located in buildings 400 m apart, a single $^{87}$Rb atom is loaded in an optical dipole trap. Both atoms are synchronously excited to the state $5^2P_{3/2}|F'=0,m_F=0\rangle$ to generate atom-photon entanglement in the subsequent spontaneous decay. The single photons emitted at a wavelength of 780 nm are collected using high-NA objectives and coupled into single mode fibres leading to the QFC devices. There, they are converted to telecom wavelength ($\lambda=1517$ nm) via difference frequency generation (DFG) in a PPLN waveguide located in a Sagnac interferometer type setup. This particular configuration fully maintains the polarization quantum state of the photon. The converted photons are guided to a middle station via fibre links with lengths up to 16.5 km, where the entanglement is swapped to the atoms via a Bell state measurement (BSM). After successfully generating atom-atom entanglement, the atoms are analysed independently by a read-out pulse of which the polarization, set by a half-wave plate (HWP) and quarter-wave plate (QWP), defines the measurement setting.}
\label{fig:setup}
\end{figure*}

\section*{Quantum network link with telecom interfaces} 
\noindent Our experiment consists of two similar, independent nodes and a middle station, which all are located in different laboratories, as illustrated in Fig.~\ref{fig:setup}. The shortest possible fibre connection from Node 1 (Node 2) to the middle station equals 50 m (750 m), longer fibre links are realized by inserting additional fibres on spools. The fibre length to the middle station is denoted as $L_1$ ($L_2$), with the total link length $L=L_1+L_2$. In both nodes, a single, optically trapped $^{87}$Rb atom acts as a quantum memory \citepaper{weber2006analysis}, where a qubit is encoded in the Zeeman-substates of the $5^2\text{S}_{1/2}\ket{F=1,m_F=\pm1}$ ground state, with $m_F=+1$ and $m_F=-1$ further denoted as $\ket{\uparrow}_z$ and $\ket{\downarrow}_z$, respectively.

The experimental sequence starts by generating atom-photon entanglement in each node \citepaper{volz2006observation}. For this purpose, the atoms are prepared in the initial state $5^2S_{1/2}\ket{F=1,m_F=0}$ via optical pumping and excited to the state $5^2P_{3/2}\ket{F'=0,m_F=0}$. During the spontaneous decay back to the ground state the atomic spin state becomes entangled with the polarization of the respective emitted photon at 780 nm due to the conservation of angular momentum. This results in the entangled atom-photon state $\ket{\Psi}_{AP} = 1/\sqrt{2}(\ket{\downarrow}_z\ket{L} + \ket{\uparrow}_z\ket{R}) = 1/\sqrt{2}(\ket{\downarrow}_x\ket{V} + \ket{\uparrow}_x\ket{H})$, where $\ket{L}$ and $\ket{R}$ denote left- and right-circular photonic polarization states, $\ket{H}$ and $\ket{V}$ denote horizontal and vertical linear polarizations. A custom made high-NA objective is used to collect the atomic fluorescence into single-mode fibres.

Photons with a wavelength of 780 nm now would suffer an attenuation by a factor of 10 after propagation through 2.5 km fibre. To overcome such high attenuation loss, we employ polarization-perserving quantum frequency conversion (QFC) to transform the wavelength of the photons to the telecom S band, where one expects attenuation by a factor of 10 only after about 50 km transmission. The QFC is realized by mixing the photons with a strong pump field at 1607 nm inside a nonlinear waveguide crystal, converting the wavelength to 1517 nm via difference frequency generation. Various spectral filtering stages, including a narrow band filter cavity (27 MHz FWHM), separate the single photons from the strong pump field and the anti-stokes Raman background originating from this field. In the shortest fibre configuration this results in a background of approximately 160 and 170 cps registered at the middle station for light from Node 1 and 2, respectively. Both converters achieve an external device efficiency of 57\%. The pump light is conveniently distributed to the nodes using the telecom fibre network and hence ensures indistinguishable frequencies of the single photons after conversion. For more details about the QFC system and an analysis of the atom-photon entanglement distribution at telecom wavelength see Ref. \citepaper{leent2020long} and Methods.

After frequency conversion, the photons are guided to the middle station with fibres of different lengths where a Bell-state measurement (BSM) swaps the entanglement to the atoms \citepaper{hofmann2012heralded,rosenfeld2017event}.
The fidelity of the BSM, and hence of the entanglement swapping protocol, is determined by the photons temporal, spectral, and spatial indistinguishability \citepaper{rosenfeld2011towards}.
This is optimized by different means, first, the photons impinge on a balanced, single-mode fibre beam splitter to guarantee a perfect spatial overlap. Second, the entanglement generation process in the nodes is synchronized to $<300$ ps, which is much smaller than the coherence time of the photons determined by the lifetime of the excited state of 26.2 ns.
And third, polarization drifts in the long fibres are compensated using an automated polarization control \citepaper{rosenfeld2008towards}. 
The photons are detected with four superconducting nanowire single-photon detectors (SNSPD), which all have an efficiency of $>85\%$ and a darkcount rate of $<65$ cps.

\begin{figure*}
\includegraphics[]{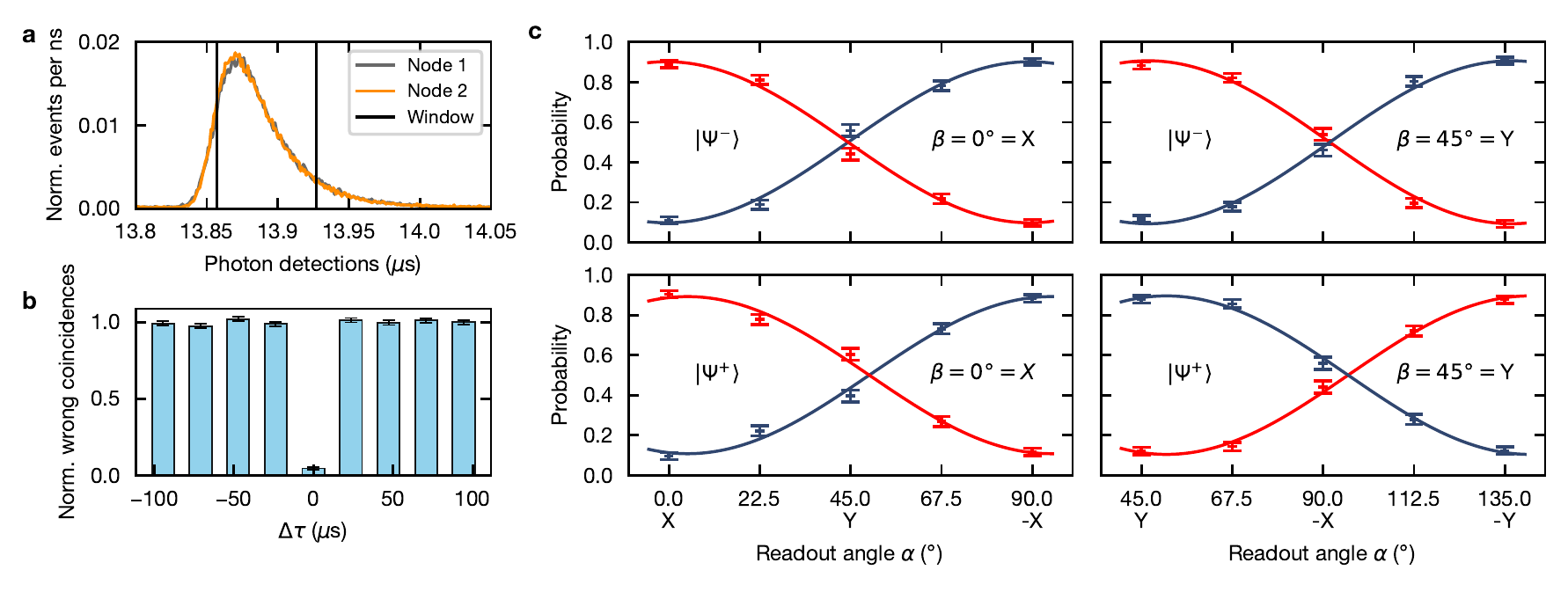} 
\caption{\protect\textbf{Characterization of the atom-atom entanglement for a fibre length of $L=6$ km.} \textbf{a}, Detection time histogram of the photons originating from Node 1 and 2 relative to the time of excitation in Node 1. For the indicated acceptance window, we observe a SBR of 58 and 65 for Node 1 and 2, respectively. The temporal overlap of the two photons is $>0.98$. \textbf{b}, Two-photon interference based on the Hong-Ou-Mandel effect. Shown is the normalized number of wrong coincidences for various time differences between the two photon wave-packets ($\Delta\tau$). \textbf{c}, Atom-atom state correlations for the $|\Psi^-\rangle$ (top) and $|\Psi^+\rangle$ (bottom) states. The correlation probability of the measurement outcome in the nodes is shown in blue, while the anti-correlation probabilities are marked in red. The data are fitted with sinusoidal functions resulting in an estimated fidelity for $|\Psi^-\rangle$ and $|\Psi^+\rangle$ relative to a maximally entangled state of 0.826(18) and 0.806(20), respectively.}
\label{fig:atom-atom-fringe}
\end{figure*}

The employed BSM setup analyses the photons in the H/V basis and hence heralds the following two Bell states

\begin{equation}
\ket{\Psi^\pm}=\frac{1}{\sqrt{2}}(\ket{\uparrow}_x\ket{\downarrow}_x\pm\ket{\downarrow}_x\ket{\uparrow}_x).
\label{eq:atomatom}
\end{equation}

\noindent Two-photon coincidences are triggered within a hard-wired 208 ns long window which sends this heralding signal back to the nodes. The signal is delayed electronically by $t\geq\ell/\frac{2}{3}c$ to simulate the signalling time back to the nodes, where $\frac{2}{3}c$ approximates the speed of light in an optical fibre and $\ell$ equals $L_1$ or $L_2$ for Node 1 or 2, respectively. Although lowering the observed final fidelity, this delay is essential to study the performance of the quantum network link in a realistic scenario.

The quantum state of the atomic memories is finally analysed with a state-selective ionization scheme, whereby the state selectivity is controlled by the polarization of a readout laser pulse \citepaper{leent2020long}. Both memories have a coherence time T$_2$ of approximately 330 $\upmu$s, which is achieved by active stabilization of magnetic fields ($<$0.5 mG) and applying a bias field of tens of milligauss along the y axis. Currently, the coherence time is limited by magnetic field fluctuations along the bias field direction and a position-dependent dephasing originating from longitudinal field components of the strongly focussed dipole trap. For more details and a simulation of the limiting decoherence mechanisms see Methods. 

\section*{Entanglement distribution at telecom wavelength}
\noindent The evaluation of the entanglement distribution over long fibre links is detailed first exemplarily for a fibre configuration of $L=6$ km. The entanglement generation rate is determined by three factors: the probability of generating an entangled state between the atoms after a synchronized excitation attempt, which is predominantly reduced by the photon collection efficiency in the nodes and amounts to $3.66 \cdot 10^{-6}$; the repetition rate of 30.8 kHz, which is mainly limited by the communication time between the nodes; and the duty cycle of approximately 1/2 which includes the fraction of time that an atom is present in both traps. This leads to an event rate of 1/19 s$^{-1}$ resulting in $N=10\,290$ entanglement generation events within 54 hours.

Facing Raman background from two QFC devices in addition to detector dark counts, a 70 ns photon acceptance window is applied during the data post-processing, as shown in Fig.~\ref{fig:atom-atom-fringe}a. This results in a signal-to-background ratio (SBR) of 58 (65) for detecting a single photon from Node 1 (Node 2), which is significantly higher than in a previously reported work (Ref. \citepaper{leent2020long}) thanks to a more favourable pump-signal frequency combination with respect to the Raman background. For the coincidence detections this leads to a SBR of 48 while accepting approximately 65\% of the recorded events. 

The quality of the two-photon interference of the converted photons is quantified by the relative occurrence of wrong detector coincidences~\citepaper{hofmann2012heralded}. These coincidences, i.e., (V$_1$,V$_2$) and (H$_1$,H$_2$), should not occur for perfectly interfering, fully-unpolarized photons. For temporally well overlapping photons ($\Delta\tau=0$), but without background correction, this results in an interference contrast of $0.955(7)$ (Fig.~\ref{fig:atom-atom-fringe}b). Here, double-excitation events in the nodes reduce the indistinguishability of the detected photons by changing the temporal shape. By rejecting early detection events, this effect is reduced at the cost of a lower event rate. For more details see Methods.

To evaluate the atom-atom entanglement we measured the atomic spin states in the two linear bases, $X$ and $Y$. For this, the polarization analysis angle in Node 2 was set to $\beta=0^{\circ}=X$ and $\beta=45^{\circ}=Y$, while the analysis angle in Node 1 was varied over $90^{\circ}$ in steps of $22.5^{\circ}$ starting from $\alpha=0^{\circ}=X$ and $\alpha=45^{\circ}=Y$, respectively. The atom in Node 1 (Node 2) was analysed at $t_1=28.5~\upmu$s ($t_2=35.5~\upmu$s) after the respective excitation pulse. The resulting atomic state correlation probabilities $P_{corr} = (N^{\alpha,\beta}_{\uparrow\uparrow}+N^{\alpha,\beta}_{\downarrow\downarrow})/N^{\alpha,\beta}$ and anti-correlation probabilities $P_{acorr} = (N^{\alpha,\beta}_{\uparrow\downarrow}+N^{\alpha,\beta}_{\downarrow\uparrow})/N^{\alpha,\beta}$ are shown in  Fig.~\ref{fig:atom-atom-fringe}c. The data are fitted with sinusoidal curves giving average visibilities of $\bar{V}=0.804(20)$ for $|\Psi^-\rangle$ and $\bar{V}=0.784(23)$ for $|\Psi^+\rangle$. To estimate the state fidelity we need to consider that the third ground state ($5^2\text{S}_{1/2}|\text{F}=1,m_F=0\rangle$) can be populated, hence operating effectively in a $3\times3$ state space. Therefore, a lower bound on the fidelity is given by $\mathcal{F}\geq\frac{1}{9}+\frac{8}{9}\bar{V}=0.826(18)$ for $|\Psi^-\rangle$ and 0.806(20) for $|\Psi^+\rangle$, relative to a maximally entangled state. Furthermore, the chosen analysis angles allow to evaluate the Clauser-Horne-Shimony-Holt (CHSH) S value \citepaper{chsh1969} for the settings $\alpha=22.5^{\circ}$, $\beta=0^{\circ}$;  $\alpha'=67.5^{\circ}$, $\beta=0^{\circ}$;  $\alpha'=67.5^{\circ}$, $\beta=45^{\circ}$; and $\alpha''=112.5^{\circ}$, $\beta'=45^{\circ}$, whereby $\alpha''$ replaces $\alpha=22.5^{\circ}$. This results in an observed value of $S=2.244(63)$, violating the limit of 2 with $3.9\sigma$. 

\section*{Entanglement distribution over up to 33 km fibre}
\begin{figure}
\includegraphics[]{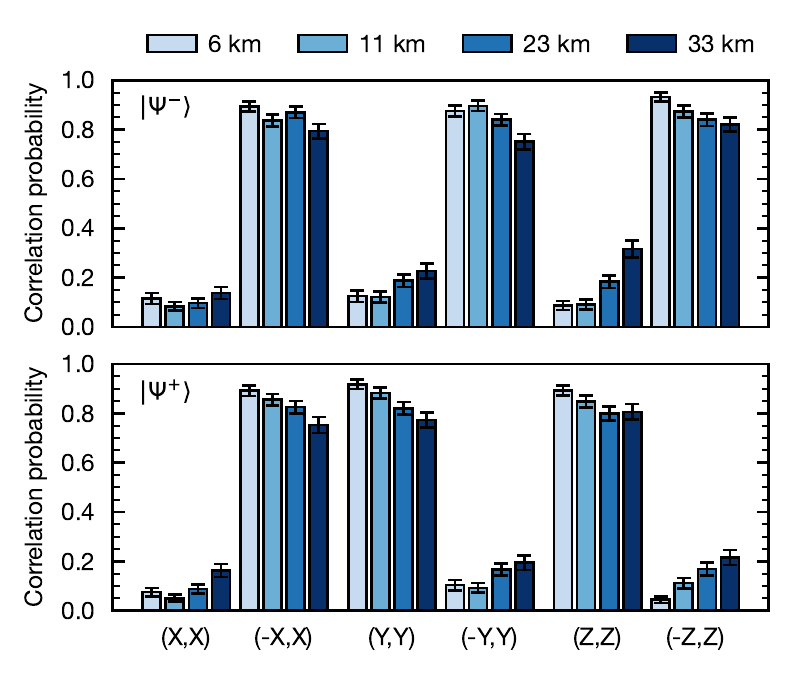} 
\caption{\protect\textbf{Observation of heralded atom-atom entanglement generated over long fibre links.} 
Correlations between the measurement results when analysing the generated atom-atom Bell states for various link lengths. For each link, the states were analysed in 3 conjugate bases (Node 1, Node 2), whereby the correlation probability of the measurement result in the nodes equals $P_{corr}=(N_{\uparrow\uparrow}+N_{\downarrow\downarrow})/(N_{\uparrow\uparrow}+N_{\uparrow\downarrow}+N_{\downarrow\uparrow}+N_{\downarrow\downarrow})$. For the different fibre lengths (short to long) 4281, 4271, 4153, and 3022 entanglement events were recorded within a measurement time of 11, 65, 97, and 175 hours. Of these events, 62-72\% was within the two-photon coincidence acceptance window, resulting in $N=185$ to 225 events per data-point. The errorbars indicate statistical errors of one standard deviation.}
\label{fig:results}
\end{figure}

\noindent To determine the effect of long fibre links, we performed a series of measurements generating and observing atom-atom entanglement in fibre configurations with a length $L$ of 6, 11, 23, and 33 km. For longer links, the event rate reduces due to both the signal attenuation of 0.22 dB/km and the longer communication times. At 33 km, this results in a repetition rate of 9.7 kHz, a success probability of $1.22\cdot 10^{-6}$, and an event rate of 1/208 s$^{-1}$. In these measurements, the entanglement fidelity relative to maximally entangled states was analysed by measurements along three bases (X, Y, and Z). Fig.~\ref{fig:results} shows the probability of correlated measurement outcomes in the nodes for each measurement setting combination and fibre configuration.

The fidelity of the observed states is estimated by first determining the contrast in the three measurement bases independently. This is done by taking the absolute difference of the two measured correlation probabilities, $E_{k}=|P_{k,k}-P_{-k,k}|$ for $k\in\{X,Y,Z\}$, from which the average contrast is computed as $\bar{E} = (E_X + E_Y + E_Z)/3$. When averaging over the observed states $\ket{\Psi^\pm}$---which show within our measurement precision similar visibilities---and again assuming the $3\times3$ state space, this results in a lower bound on the fidelities $\mathcal{F}=$ 0.830(10), 0.799(11), 0.719(12), and 0.622(15) relative to maximally entangled states for $L$ equals 6, 11, 23, and 33 km, respectively. The estimated fidelity for the 6 km fibre configuration is in good agreement with the fidelity estimated from the fringe measurements in two bases presented before. Moreover, all observed fidelities clearly exceed the bound of 0.5 and hence witness an entangled state. 

\begin{figure}
\includegraphics[]{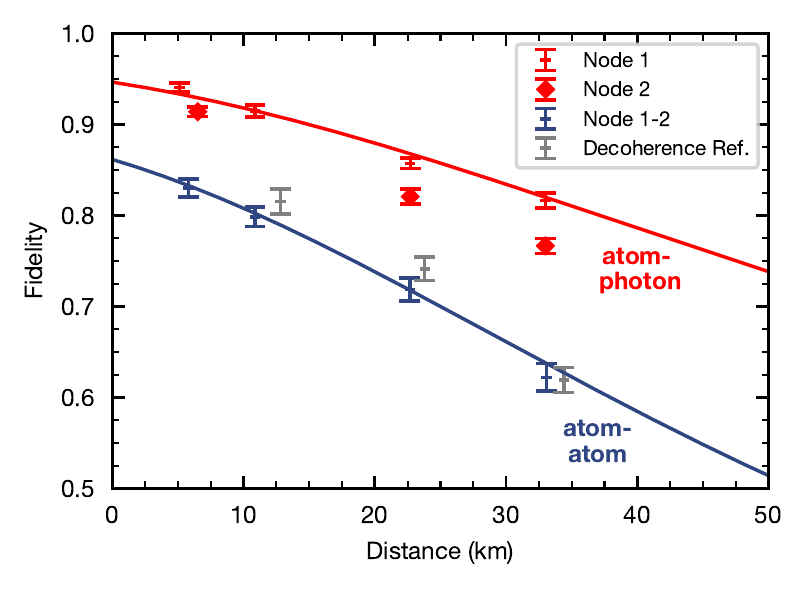}
\caption{\protect\textbf{Observed entanglement fidelity for various link lengths.} Overview of the observed atom-atom fidelities for different fibre configurations (blue). For completeness, the observed atom-photon fidelities of the states between the nodes and the middle station (red), with $L=2 \cdot L_{i}$, are given (see Methods). The solid lines are simulations based on a model taking into account the decoherence of the atomic memories, see Methods. The gray points are reference measurements of the atom-atom state decoherence, without long fibres, but with corresponding readout delay.}
\label{fig:overview}
\end{figure}

The observed fidelities are shown in dependence of the fibre link length in Fig.~\ref{fig:overview}, where also the measured atom-photon entanglement fidelities for states shared between the nodes and the middle station are shown for completeness. 
The atom-atom state fidelity for different fibre lengths $\mathcal{F}(L)$ is modelled based on simulations of the generation and evolution of the two atom-photon states, which are shown with solid lines. 
We estimate the visibility of the atom-atom state by the product of the two atom-photon visibilities and the interference contrast of the BSM \citepaper{zukowski1995entangling}. 
Evidently, decoherence of the atomic states dominates the loss in fidelity for longer fibre links. 
For $L=33$ km, the states were analysed 171 $\upmu$s and 178 $\upmu$s after excitation in Node 1 and 2, respectively, which approaches the coherence time of the states. 
In contrast, the SBR is robust to an increase in fibre length since both the single photons as well as the QFC background are attenuated in the long links. A minor reduction in SBR (42 for $L=33$ km) is explained by relatively more detector dark counts and can be solved by installing detectors with lower dark counts. Also, polarization drifts are comparably well compensated in all link configurations, see Methods. 

To verify that the memory decoherence limits the loss in fidelity for long fibre links, we performed a series of measurements without additional fibres inserted, however, with the memory readout times electronically delayed according to the long fibre links. The observed fidelities are shown in gray in Fig.~\ref{fig:overview} at $L=\frac{2}{3}c(t_1+t_2)/2$ (matching the two-way communication time to the middle station for distance $L$) and show, within the measurement accuracy, no difference in observed fidelity compared to the configuration with long fibres. 

\section*{Discussion and Outlook}
\noindent The results clearly indicate the feasibility of turning to large-scale quantum networks by facilitating an increase in line-of-sight separation of the nodes to tens of kilometres.
Possible improvements include increasing the coherence time of the atomic states by implementing a new trap geometry to mitigate the position-dependent dephasing in combination with a state-transfer to a qubit encoding less sensitive to magnetic fields \citepaper{korber2018decoherence}. This will allow to generate entanglement of quantum memories on a suburban scale with a high fidelity.

To conclude, employing efficient telecom interfaces in our nodes enabled the generation of heralded entanglement between two atomic quantum memories over fibre links with a length up to 33 km. The analysis of the results clearly shows that improvements on the memory coherence time are mandatory but will allow to entangle two atomic quantum memories with a fidelity better than 80\% over fibre lengths up to 100 km, thereby paving the way towards long-distance entanglement distribution for future quantum repeater networks. \\

\bibliographystylepaper{unsrt}
\bibliographypaper{atom-atom_telecom_bib}


\newpage

\section*{Methods}

\noindent\textbf{Atom-photon entanglement distribution at telecom wavelength} \\
The polarization-preserving quantum frequency conversion (QFC) devices used in this work are described in detail in references \citepaper{leent2020long} and  \cite{bock2021thesis}. In contrast to previous work, here a more favourable pump--signal frequency combination is selected with respect to the Raman background: 1607 nm--1517 nm instead of 1600 nm--1522 nm. This increases the signal-to-background ratio (SBR) by a factor of 4 and allows to install a QFC device in Node 2---effectively doubling the background---while not being limited by the SBR.

Since the quality of the entanglement shared between the two nodes directly depends on the fidelity of the two entangled atom-photon pairs, we individually characterize the atom-photon entanglement generated in both nodes. The generated states are analysed using the same fibre configurations and atomic readout times as during the atom-atom entanglement measurements presented in the main text. For an overview see Table \ref{tab:config}. Note that an high fidelity atomic state readout can only be made after a full oscillation period of the atom in the dipole trap, which equals 14.3 $\upmu$s and 17.8 $\upmu$s for Node 1 and 2, respectively.

\begin{table}[h]
\caption{\textbf{Long fibre configurations and corresponding atomic readout times.} The fibre link lengths and corresponding atomic readout times for the experiments presented in the main text and the atom-photon characterization measurements. $L_1$ ($L_2$) equals the fibre length between Node 1 (Node 2) and the middle station. $A_1$ ($A_2$) gives the attenuation in the fibre network between the node and the middle station, this includes inefficiencies of fibre-to-fibre connectors.}
\bgroup
\def\arraystretch{1.3}%
\begin{tabular}{ c || c | c | c | c | c | c  }
$L$ (km) & $L_1$ (km)  & $L_2$ (km) & $A_1$ (dB) & $A_2$ (dB) & $t_1$ ($\mu$s) & $t_2$ ($\mu$s) \\ \hline 
6  & 2.6  & 3.3 & -0.7 & -0.8 & 28.5  & 35.5 \\
11 & 5.4  & 5.5 & -1.5 & -1.3 & 57.1  & 71.0 \\
23 & 11.3 & 11.4 & -3.3 & -2.8 & 114.2 & 124.3  \\
33 & 16.5 & 16.6 & -4.5 & -4.1 & 171.2 & 177.5  \\ 
\end{tabular}
\egroup
\label{tab:config}
\end{table}

\begin{figure}
\includegraphics[width=0.99\linewidth]{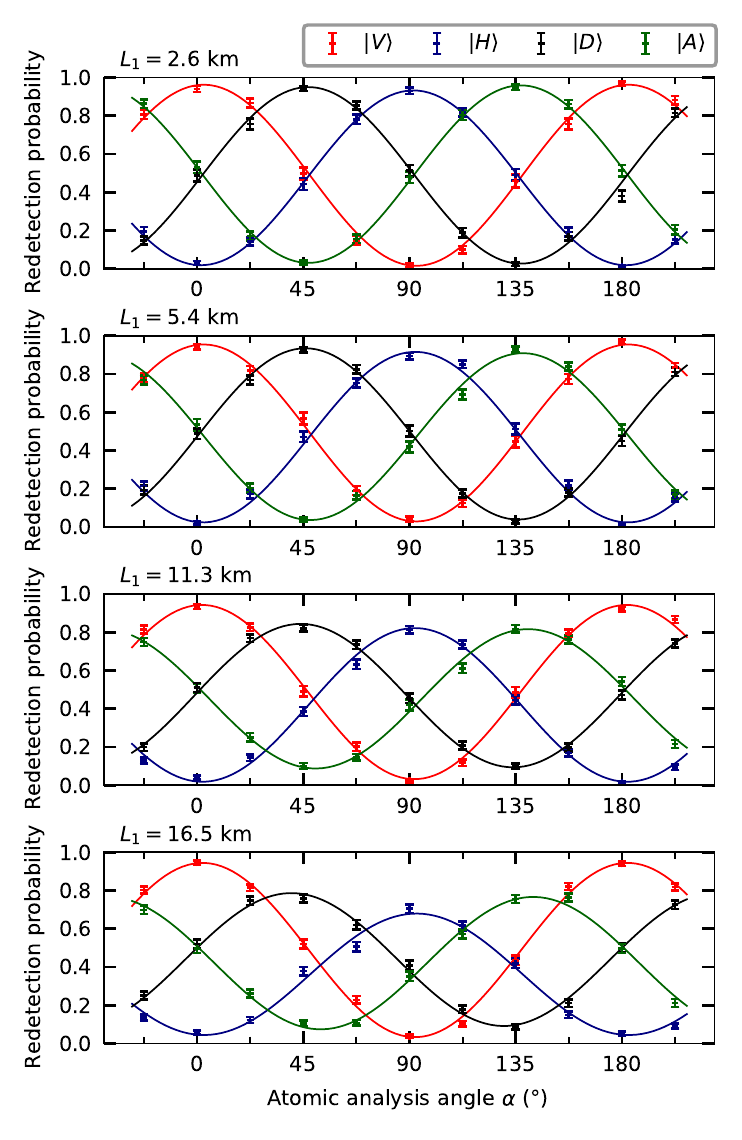}
\caption{\protect\textbf{Atom-photon entanglement distribution for Node 1.} The measurements include an atomic readout delay to allow for two-way communication with the middle station. The error bars indicate statistical errors of one standard deviation.}
\label{fig:atom1-photon}
\end{figure}

\begin{figure}
\includegraphics[width=0.99\linewidth]{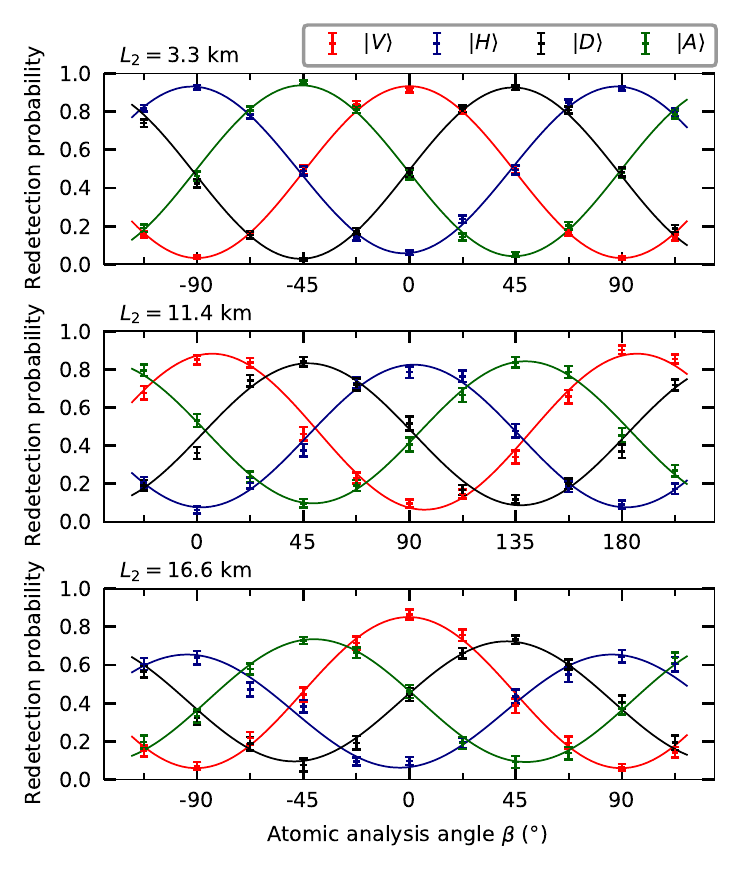}
\caption{\protect\textbf{Atom-photon entanglement distribution for Node 2.} The measurements include an atomic readout delay to allow for two-way communication with the middle station.}
\label{fig:atom2-photon}
\end{figure}

The atom-photon entanglement fidelity is analysed following the methods in Ref. \citepaper{leent2020long}, whereby the atomic readout time is now delayed to allow for two-way communication to the middle station for each node over the respective fibre length. The polarization of the photons are measured in two bases, $H/V$ (horizontal/vertical) and $D/A$ (diagonal/anti-diagonal), i.e., X and Y, while the atomic analysis angle was rotated over angles including these bases. The atom-photon state correlations are shown for Node 1 in Fig. \ref{fig:atom1-photon} and for Node 2 in Fig. \ref{fig:atom2-photon}. 

For the fibre configuration L = 6 km, i.e., $L_1=2.6$ km and $L_2=3.3$ km, we find atom-photon state fidelities of 0.941(5) for Node 1 and 0.911(6) for Node 2, relative to a maximally entangled state, which are mainly limited by the atomic state readout and entanglement generation fidelity. For longer fibre lengths the entangled state decoherence due to magnetic field fluctuations along the guiding field and the position-dependent dephasing. \\

\noindent\textbf{Modelling of the quantum memory decoherence} \\
In both nodes, a single $^{87}$Rb atom is stored in an optical dipole trap, where a qubit is encoded into the states $5^2\text{S}_{1/2}\ket{F=1,m_F=\pm1}$. The dipole trap is operated at $\lambda_{\text{ODT}}=850$ nm with typical trap parameters of, e.g., for Node 1, a trap depth $U_0=2.32$ mK and beam waist $\omega_0=2.05 \upmu$m. 
The qubit evolves effectively in a spin-1 system since the state $5^2\text{S}_{1/2}\ket{F=1,m_F=0}$ could also be populated. The state fidelity is influenced by two factors: the first one is the AC-stark shift originating from the dipole trap and, secondly, the Zeeman effect arising from magnetic fields.

To model the dephasing of the quantum memories, we simulate the evolution of this spin-1 system while the atom is oscillating in the dipole trap, affected by longitudinal polarization components and external magnetic fields \cite{PhD-Daniel}. For this, first, we randomly draw a starting position and velocity of an atom from a 3D harmonic oscillator distribution in thermal equilibrium. Second, the motion of the atom is simulated in a realistic Gaussian potential resulting in an atomic trajectory for which the evolution of the atomic state is calculated based on the local optically induced and external magnetic fields. Finally, this is repeated for a large number of trajectories, whereby the averaged projection for all trajectories yields the simulation result.

\begin{figure*}
\includegraphics[]{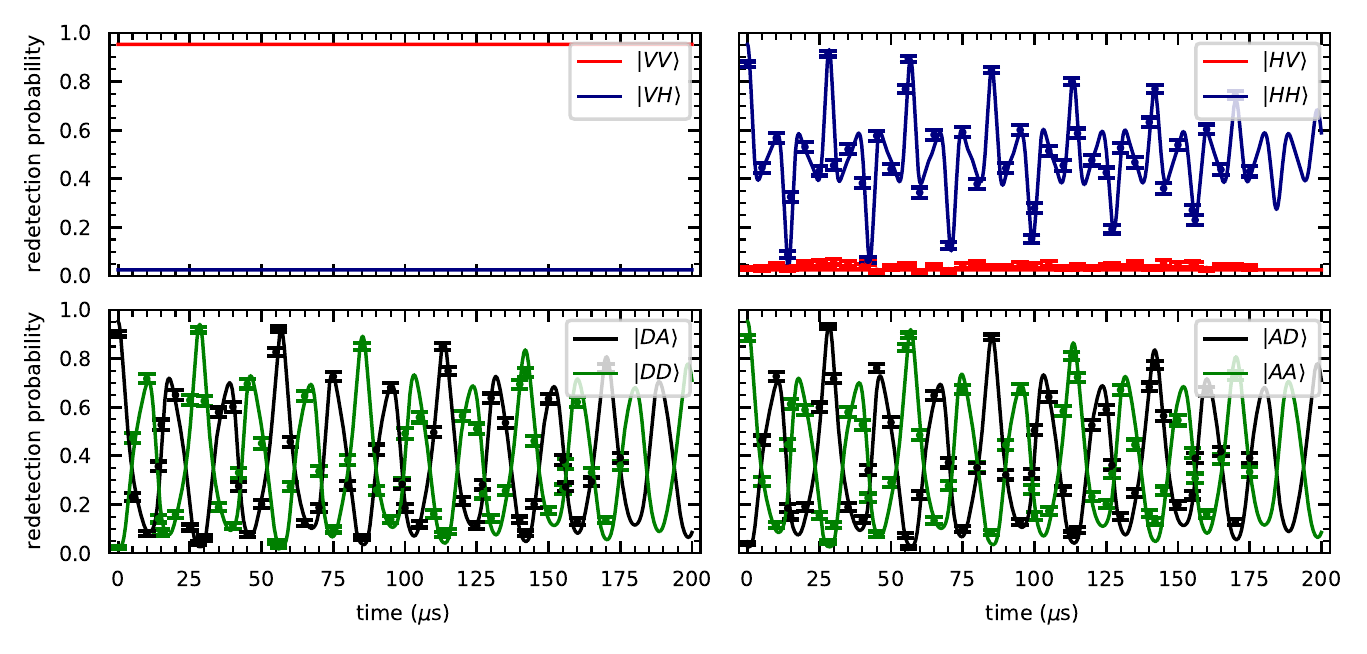}
\caption{\protect\textbf{Atomic state evolution in two bases.} Atom-photon entanglement simulations (solid lines) and measurements (points) of Node 1 with photon detection at 780 nm using a 5 m long fibre. The atomic state readout orientation and time is varied to characterize the memory coherence. The labels in the legend indicate the $|atom\,\,photon\rangle$ state analysed. Simulation parameters: trap waist $2.05 \upmu$m, trap depth $2.32$ mK, atom temperature $50\upmu K$, bias field $B_y=75.5$ mG, and field fluctuations $\Delta B_y=0.5$ mG Gaussian distributed. With these parameters, we observe de- and rephasing of the atomic state due to the longitudinal polarization components at the trap frequency of 70 kHz and a larmor precession at a frequency of 105 kHz.}
\label{fig:coherence}
\end{figure*}

The model takes the following independently measured inputs: (1) the trap geometry specified by the beam waist $\omega_0$, which is obtained from knife-edge measurements of the dipole trap beam focus in two dimensions \cite{PhD-Robert}; (2) the trap depth $U_0$, determined via measurements of the transverse trap frequency using parametric heating \cite{alt2003single} and the atomic state rephasing period \cite{PhD-Daniel}; and (3) the atomic temperature $T$, modelled as a Boltzmann distribution which is measured via the release and recapture technique \cite{tuchendler2008energy}. Inputs 1 and 2 define the position, amplitude, and phase of the longitudinal polarization components, while inputs 1-3 characterize the atomic trajectories. Furthermore, we include a uniform magnetic field along three directions with additional shot-to-shot noise following Gaussian distributions. 

Fig.~\ref{fig:coherence} shows simulation results and measurement data of the state evolution in Node 1 for varying state readout orientation and time. The model accurately predicts the evolution of the measured atomic states and shows that the memory storage time is limited by magnetic field fluctuations on the order of $<0.5$ mG along the bias field direction in addition to the position-dependent dephasing due to the longitudinal field components of the strongly focussed dipole trap. The simulation results presented in the main text consider the envelope of the found oscillating state evolution in three bases. \\

\begin{figure*}
\includegraphics[width=0.7\linewidth]{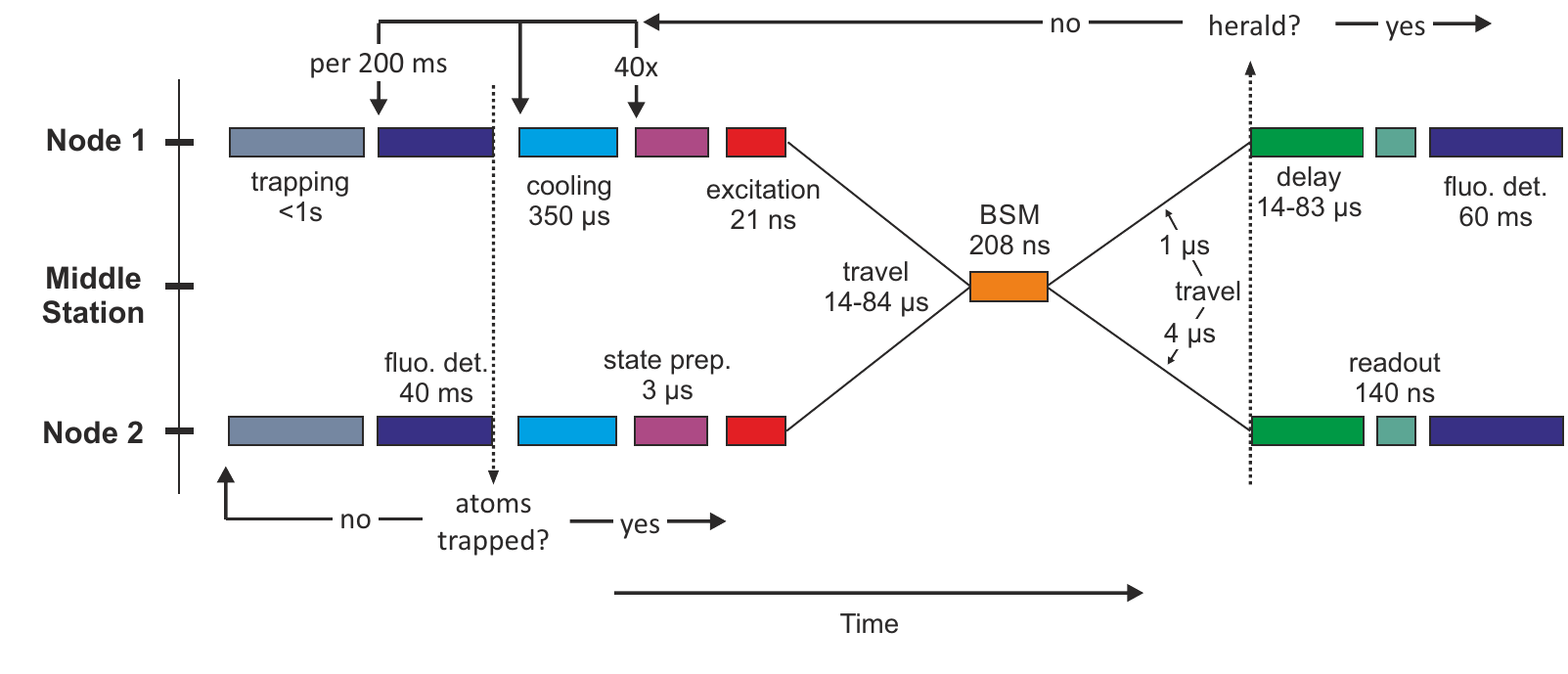}
\caption{\protect\textbf{Experimental sequence of the atom-atom entanglement generation.} In both nodes a single atom is trapped and cooled using polarization gradient cooling. Next, the entanglement generation tries start containing state preparation and synchronous excitation of the atoms. The atoms are re-cooled after 40 unsuccessful entanglement generation tries. The travel time of the photons to the middle station equals 14 to 84 $\upmu$s, depending on the fibre length. With regard to the assumption that the heralding signal is communicated back to the nodes along fibres of respective lengths, an additional delay is included to account for the communication time from the middle station back to the nodes. After a 200 ms interval of entanglement generation tries the presence of the atoms in the traps is checked using fluorescence collection with an APD (see text). The QFC takes place subsequent to the atomic excitation.}
\label{fig:sequence}
\end{figure*}

\noindent\textbf{Experimental sequence} \\
The entanglement generation sequence is visualised in Fig.\ref{fig:sequence}. The sequence starts by trapping an atom in both nodes. For this, a single atom is loaded from a magneto-optical trap (MOT) into a tightly focussed dipole trap, which takes less than 1 second. 
Every entanglement generation try consist of 3 $\upmu$s optical pumping (80\% efficiency) and an excitation pulse (Gaussian laser pulse with a FWHM of 21 ns) to generate atom-photon entanglement in the following decay. Subsequent to each try, a waiting time is implemented to cover the propagation time of the photons in the long fibres. After 40 unsuccessful tries the atoms are cooled for 350 $\upmu$s using polarization gradient cooling (PGC). 
The lifetime of the atoms in the trap during the entanglement generation process is approximately 5 seconds.

To verify if both traps still store a single atom, the entanglement generation process is interrupted after 200 ms to check the presence of the atoms. For this, a microelectromechanical systems (MEMS) fibre-optic switch is implemented in each node at the SM-fibre originating from the atom trap. The switches guide the atomic fluorescence either to the QFC devices during the entanglement generation tries or to an avalanche photodiode (APD) located at each node during 40 ms of fluorescence collection. Note that the SNSPDs of the BSM cannot be used for this purpose since they are behind narrowband spectral filters. \\

\noindent\textbf{Polarization control of long fibres} \\
Polarization drifts in the long fibres are compensated for using an automated polarization control based on the method presented in Ref. \citepaper{rosenfeld2008towards}. 
The polarization control is performed every 7 minutes, takes approximately 20 seconds, and is based on a gradient decent optimization algorithm.
In this way, polarization errors are kept below $1$\% during all measurements.

The fibre polarization is optimized using classical laser light at the single photon frequency whereby the polarization is alternated between vertical and diagonal linear polarizations at 10 Hz. The light is overlapped with the complete single photon path up to the detectors originating from Node 1 and Node 2. In both output arms of the BS a flip-mirror can reflect the classical light into a polarimeter during the optimization. Three fibre polarization controllers (PC) are connected to the fibre BS of the BSM: at both input ports and at one output port. \\

\noindent\textbf{Entanglement swapping fidelity} \\
The single photons are detected with a Bell-state measurement device consisting of a fibre beamsplitter (BS), two polarizing beamsplitters (PBS) (Wollaston Prisms), and four superconducting nanowire single photon detectors (SNSPD), as illustrated in Fig. 1 of the main text.
In this setup, the fibre BS guarantees a unitary spatial overlap of the photons originating from the nodes, while the PBSs and single photon detectors allow for polarization analysis in both output ports.
The detectors, labelled H$_1$, V$_1$, H$_2$, and V$_2$, are not photon number resolving, and hence coincidences events in six detector pairs can be registered, see Tab.~\ref{tab:detections}.
For the purpose of a Bell-state measurement, we can categorize these combinations into three groups: $D_+$, $D_-$, and $D_\varnothing$. 
Here, detector combinations in group $D_+$ and $D_-$ herald the Bell states $|\Psi^+\rangle$ and $|\Psi^-\rangle$, respectively, while combinations in group $D_\varnothing$ should not occur for perfectly interfering photons and are discarded in the analysis. 
However, the relative occurrence of these events is used in the following to quantify the two-photon interference contrast.

For not interfering photons, two-photon events are evenly distributed between the 16 possible detector combinations (not considering experimental imperfections). Since the order of the detector combination is not of interest, e.g., (V$_1$,H$_1$) is similar to (H$_1$,V$_1$), we end up with 10 distinct coincidences and their probabilities, as listed in Tab.~\ref{tab:detections}. For perfectly interfering photons the probabilities differ: the probability to detect the $D_\varnothing$ group vanishes and all four Bell states are detected with a probability of 1/4, whereby the $|\Phi^{\pm}\rangle$ Bell states fall into the group 'not detected' for the employed setup.

The two-photon interference contrast is defined as \citepaper{hofmann2012heralded}

\begin{equation}
C=1-\frac{2N_{D_\varnothing}}{N_{D_+} + N_{D_-}}.
\label{eq:contrast}
\end{equation}

\noindent where $N_k$ is the number of events in detection group $k$. With this definition and the probabilities of the different coincidences, the contrast equals zero for not interfering photons and one for perfectly interfering photons. See Ref. \cite{hofmann2014thesis} for a thorough analysis of the two-photon interference contrast and entanglement swapping fidelity, including experimental imperfections.

The interference contrast is measured as follows. During measurement runs all single-photon detection events are recorded, which allows to count the number of occurrences of the coincidence events for all three detection groups. Next, the the interference contrast is evaluated using equation (\ref{eq:contrast}). To verify this method, we additionally evaluate the contrast of not interfering photons. This is done by analysing coincidence detections of two photons originating from distinct entanglement generation tries. In this way, the photons did not interfere since the photon wave-packages are completely separated in time. Fig. 2b of the main text shows exactly this for the $L=6$ km measurement. Shown are the normalized wrong coincidences, defined as $1-C$, for varying time differences between the photons ($\Delta\tau$). Note that the horizontal spacing of the measurement times equals the repetition rate of the entanglement generation tries.

The entanglement swapping fidelity is mainly limited by two effects. First, by experimental imperfections that reduce the indistinguishability of the two photons, e.g., as discussed in the main text, by a not perfect time overlap of the two photon wave-packets. Second, by double excitations due to the finite duration of the excitation pulse. For a detail description see Ref. \cite{zhang2021experimental}.

\begin{table*}
\caption{\textbf{Possible two-photon coincidences.} The probabilities of the events are given for not interfering photons and for perfectly interfering photons.}
\bgroup
\def\arraystretch{1.3}%
\begin{tabular}{ c | c || c | c   }
detection & coincidence & no interference & perfect interference  \\ 
\hline \hline
\multirow{4}{8em}{not detected} 	& H$_1$,H$_1$ & 1/16 & 1/8 \\ \cline{2-4}
						& H$_2$,H$_2$ & 1/16 & 1/8 \\ \cline{2-4}
						& V$_1$,V$_1$ & 1/16 & 1/8 \\ \cline{2-4}
						& V$_2$,V$_2$ & 1/16 & 1/8 \\ 
\hline 
\multirow{2}{8em}{D$_\varnothing\rightarrow$ discarded} 	& H$_1$,H$_2$ & 1/8 & 0 \\ \cline{2-4}
						& V$_1$,V$_2$ & 1/8 & 0 \\ 
\hline 
\multirow{2}{8em}{D$_+\rightarrow|\Psi^+\rangle$} 	& H$_1$,V$_1$ & 1/8 & 1/8 \\ \cline{2-4}
						& H$_2$,V$_2$ & 1/8 & 1/8 \\ 
\hline 
\multirow{2}{8em}{D$_-\rightarrow|\Psi^-\rangle$} 	& H$_1$,V$_2$ & 1/8 & 1/8 \\ \cline{2-4}
						& V$_1$,H$_2$ & 1/8 & 1/8 \\
\hline

\end{tabular}
\egroup
\label{tab:detections}
\end{table*}

\bibliographystyle{unsrt}
\bibliography{atom-atom_telecom_bib}

\section*{Acknowledgements}
\noindent M.B. acknowledges the hospitality of the LMU group during multiple stays. We acknowledge helpful discussions with Stephan Kucera. We acknowledge funding by the German Federal Ministry of Education and Research (Bundesministerium f{\"u}r Bildung und Forschung (BMBF)) within the projects Q.com.Q and Q.Link.X (Contracts No. 16KIS0127, 16KIS0123, 16KIS0864, and 16KIS0880), the Deutsche Forschungsgemeinschaft (DFG, German Research Foundation) under Germany’s Excellence Strategy – EXC-2111 – 390814868, and the Alexander von Humboldt foundation.

\end{document}